\documentclass[12pt]{iopart} 

\usepackage{epsfig}
\usepackage{graphics}

\begin{document}

\title{What makes slow samples slow in the Sherrington-Kirkpatrick
  model} \author{Alain Billoire} \address{ Institut de physique
  th\'{e}orique, CEA Saclay and CNRS, 91191 Gif-sur-Yvette, France}
\date{ \today}
\begin{abstract}
Using results of a Monte Carlo simulation of the
Sherrington-Kirkpatrick model, we try to characterise the slow
disorder samples, namely we analyse visually the correlation between
the relaxation time for a given disorder sample $J$ with several
observables of the system for the same disorder sample. For
temperatures below $T_c$ but not too low, fast samples (small
relaxation times) are clearly correlated with a small value of the
largest eigenvalue of the coupling matrix, a large value of the site
averaged local field probability distribution at the origin, or a
small value of the squared overlap $<q^2>$. Within our limited data,
the correlation remains as the system size increases but becomes less
clear as the temperature is decreased (the correlation with $<q^2>$
is more robust) .  There is a strong correlation between the values of
the relaxation time for two distinct values of the temperature, but
this correlation decreases as the system size is increased. This may
indicate the onset of temperature chaos.
\end{abstract}

\date{\today}

\maketitle

The Sherrington-Kirkpatrick (SK) model has been intensively studied
since the mid-seventies, when it was introduced~\cite{SK} as a
starting point for studying spin glasses. It has a low temperature
spin glass phase, with a very slow dynamics and an equilibrium
relaxation time that diverges when the number of sites $N$ goes to
infinity. The value of the equilibrium relaxation time $\tau$
depends strongly on the disorder sample $J$, and $\tau$ is not self
averaging (namely $\Bigl[E(\tau^2)-E(\tau)^2\Bigr]/E(\tau)^2$, where
$E(\cdot)$ is the disorder average, does not go to zero as $N\to \infty$).
There are now reasonable evidences, both analytical and numerical
(see~\cite{ABEM,AB} and references therein) that, in the low
temperature phase of the model, the disorder average of the logarithm
of the equilibrium relaxation time $E(\ln{\tau})$ grows like $N^{1/3}$
as $N$ grows, and that the probability density function of
$\ln{\tau}$ scales according to the equation
$P(\ln{\tau})=N^{-\psi}F((\ln{\tau}-E(\ln{\tau}))/N^{\psi})$, with some
$N$ independent function $F(\cdot)$ with zero mean, and $\psi=1/3$,
although it has been argued\cite{MONGAR,JANKE} that $\psi$ may be
slightly less than $1/3$ (In such a case $\ln{\tau}$ would be weakly
self averaging).

Whether $\psi$ is equal to $1/3$ or slightly less than $1/3$, there
are definitively disorder samples with extremely slow dynamics. Our
aim in this note is to try to characterise these ``slow samples".  We
have two motivations in mind, one theoretical and the other more down
to earth.  The first motivation is the question of the behaviour of
$F(x)$ for large values of the argument $x$. It has been argued
in~\cite{MONGAR} that $F(x)$ has an exponential behaviour with $\ln
F(x)\propto -x^{\eta}$ for large $x$.  If furthermore the tail of
$F(x)$ is dominated by rare disorder samples with small probability
$\propto \exp(-A N^{\alpha})$ and an anomalously large free energy
barrier $\propto N^{\beta}$ (with $\beta>1/3$), the exponents
$\alpha$, $\beta$, $\psi$ and $\eta$ fulfil~\cite{MONGAR} the
consistency relation $(\beta-\psi)\eta=\alpha$. For example, a
dominance by the rare samples with all exchange couplings $J_{i,j}$
positive (up to a gauge transformation, see later) would correspond to
$\alpha=2$, and $\beta=3/2$. The value $\psi=1/3$ would then imply
that $\eta=12/7$.  This value for $\eta$ is however not compatible
with the numerical results of~\cite{MONGAR} for the distribution
$P(\log{\tau})$.  In this reference arguments are given for the values
$\alpha=\beta=1$, $\psi=1/4$ (and accordingly $\eta=4/3$) instead.
The second motivation for the characterisation of the slow samples is
of practical matter for Monte Carlo simulations: The vast majority of
the Monte Carlo simulations of disordered systems in the literature
use the same number of iterations for all disorder
samples~\footnote{For counter examples where the CPU time is adjusted
  to the disorder sample sluggishness see~\cite{BBJ} or~\cite{VMM}.}.
There is accordingly a danger that some rare slow disorder samples are
not thermalized.  One is led naturally to the idea of concentrating
the computational effort on the hard samples, that may require orders
of magnitude more iterations than the mainstream disorder samples.
Since the measurement of the relaxation time for every disorder sample
is very time consuming, any heuristic method to pinpoint the slow
samples can be valuable.

The numerical method used is similar to the one used in~\cite{AB}.  We
consider the SK model with binary exchange couplings, and the Hamiltonian

\begin{equation}
{\cal H}=-\frac{1}{2} \sum_{i\neq j} J_{i,j} \sigma_i \sigma_j \ ,
\end{equation}
with $ \sigma_i=\pm 1$, $J_{i,j} =\pm 1/\sqrt{N}$.
 We measure the dynamic overlap

\begin{equation}
q_d(t)=\frac{1}{N} \sum_{i=1}^N \sigma_i(t_0) \sigma_i(t+t_0) \ ,
\end{equation}
averaged along a very long trajectory (namely averaging over many
values of $t_0$), with the Metropolis dynamics, starting from a well
equilibrated spin configuration (obtained with the parallel tempering
algorithm).  Obviously $q_d(0)=1$ and $q_d(t)$ decreases continuously
towards zero, as $t$ grows.  We define the relaxation time $\tau$ by
the equation $q_d(t=\tau)=1/2 \sqrt{E(<q^2>)}$, where $q$ is the usual
($J$ dependent) overlap between two replica. Here we depart
from~\cite{AB} where a definition involving $<q^2>$, with no disorder
average, was used. Indeed, since we are looking for correlations
between the value of $\ln{\tau}$ and other $J$ dependent quantities
(including $<q^2>$ itself), it seems more appropriate to define $\tau$
using a sample independent condition.  The parameters of the
simulations are such that our estimates of $\ln{\tau}$ have negligible
thermal noise, as compared to the disorder sample to disorder sample
fluctuations (see~\cite{AB} for details).  That the thermal noise is
tamed is an essential condition for our analysis.  We have data for
$N=64$ to $512$,  $1024$ disorder samples, and temperatures $T=0.4,
0.5, \ldots$ (The critical temperature is $T_c=1$).

Note that as in~\cite{AB} we do not measure relaxation times larger
than some  $t_{window}\approx 3.7 \ 10^6$ (and $t_{window}\approx 1.7
\ 10^6$ for $N=512$.). For a given couple $N$ and $T$, the disorder
samples with relaxation time larger than $t_{window}$ are thus
``censored".  By convention, $\ln{\tau}=-1$ for such samples, namely
minus one means overflow.  Note that the real $\ln{\tau}$ is never
equal to minus one (our relaxation times are integers).

\begin{figure}[tbhf]
\centering
\includegraphics[width=0.3\textwidth,angle=270]{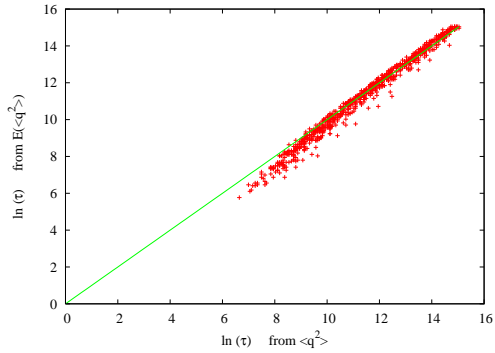}
\caption{Plot of $\ln{\tau}$, as measured using the definition of
  this note, as a function of the value obtained using the definition
  used in~\cite{AB}. Here $N=128$ and $T=0.4$.  The straight line is
  the diagonal. There are no disorder samples with $\tau>t_{window}$
  in the data.}
\label{Test}
\end{figure}

\begin{figure}[tbhf]
\centering \includegraphics[width=0.3\textwidth,angle=270]{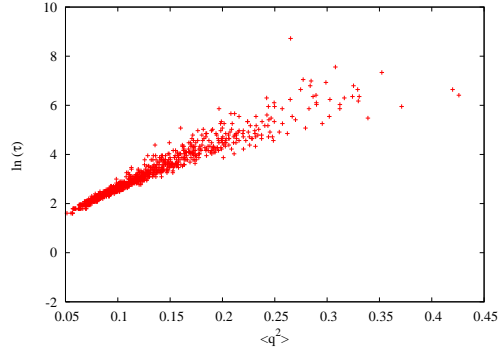}
\caption{The logarithm of the relaxation time as a function of
  $<q^2>$ for $N=64$ and $T=0.8$. There are no disorder samples with
  $\tau>t_{window}$ in the data.}
\label{Q2}
\end{figure}

As explained just before, we found it more proper to define the
relaxation time using a condition involving the disorder averaged
$E(<q^2>)$ rather than $<q^2>$ as was done in~\cite{AB}. It turns
out however that this makes little difference as shown in
figure~\ref{Test} where we compare the two definitions in the case
$N=128$, $T=0.4$.  The same conclusion holds for other couples of $N$
and $T$.  On close look, one notices that for low values of
$\ln{\tau}$ the definition used here gives systematically lower
results than the one used in~\cite{AB}.  This is explained by the fact
that low values of $\ln{\tau}$ are strongly correlated to low values
of $<q^2>$ (see later in the text and figure~\ref{Q2}).  This small
systematic difference between the two definitions of the relaxation
time disappears when $\ln{\tau}$ grows.

\begin{figure}[tbhf]
\centering
\includegraphics[width=0.3\textwidth,angle=270]{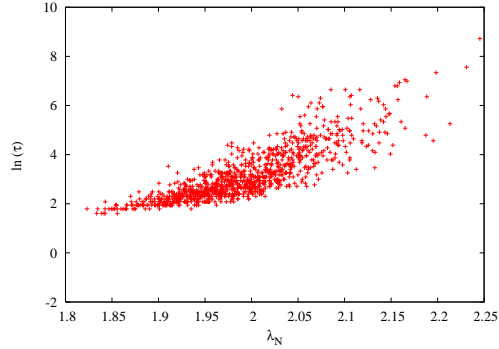}
\caption{The logarithm of the relaxation time as a function of the
  largest eigenvalue $\lambda_N$ of the $\{ J_{i,j} \}$ matrix for
  $N=64$ and $T=0.8$. There are no disorder samples with
  $\tau>t_{window}$ in the data.}
\label{eigen}
\end{figure}

The first question we would like to address is whether the tail of the
distribution $P(\ln{\tau})$ is dominated by ferromagnetic disorder
configurations. In order to proceed, we need a measure of the
ferromagnetic character of a disorder configuration.  The sum
$\sum_{i\neq j} J_{i,j}$ is not a suitable indicator since it is not
invariant under the local gauge symmetry of the model $\sigma_i \to
\epsilon_i \sigma_i$, $J_{i,j} \to \epsilon_i \epsilon_i J_{i,j}$ with
$\epsilon_i=\pm1$, whereas the dynamics is invariant under this
symmetry. Said another way, for every disorder configuration with all
$J_{i,j}>0$, there is a huge number of $2^N-1$ other gauge transformed
configurations with the same value of $\ln{\tau}$ (and the same
weight) but a different value of $\sum_{i\neq j} J_{i,j}$, and any
correlation is washed out.  (We nevertheless checked that there is
indeed no sign of correlations between $\ln{\tau}$ and $\sum_{i\neq j}
J_{i,j}$ in our data).  A better indicator of the ferromagnetic
character of a disorder configuration, that has been proposed
in~\cite{JANKE}, is the largest eigenvalue $\lambda_N$ of the matrix
$\{J_{i,j}\}$.  In the SK model the diagonal elements of this matrix
are not used, but they are obviously needed however in order to
compute the eigenvalues of the matrix, and we have set them equal to
zero. Our results for the correlation between $\ln{\tau}$ and
$\lambda_N$ can be found in figure~\ref{eigen} for $N=64$ and
$T=0.8$. The points on the $x$ axis are concentrated around the value
two, in agreement with random matrix results for the GOE with the
normalisation $E(J_{i,i}^2)=1/N$ for $i\neq j$ (in the GOE the
diagonal elements are random with $E(J_{i,i}^2)=2/N$, and not
identically zero, but this should not change the asymptotic
behaviour). There is a clear correlation between low values of
$\lambda_N$ and low values for $\ln{\tau}$, but this correlation
becomes fuzzy as $\lambda_N$ grows.  We were looking for a
characterisation of the slow disorder samples but we found a
characterisation of the fast disorder samples instead.

The scatter plots become progressively harder to interpret as $N$
grows and / or $T$ decreases, since we are missing more and more
points that correspond to censored values of $\ln{\tau}$. One can
nevertheless conclude from our data that the correlation remains as
the system size increases, but becomes more fuzzy as temperature is
decreased.  (The neat correlation seen in the left of
figure~\ref{eigen} fades away as the temperature is decreased).  The
net conclusion is that there is in our data no visible dominance of
the large relaxation region by ferromagnetic disorder
samples~\footnote{Although one may always argue that truly
  ferromagnetic disorder samples correspond to $\lambda_N$ of order
  $\sqrt{N}$, deep in the tail of the distribution that we do not see
  due to limited statistics.}. This is in agreement with the findings
of~\cite{MONGAR} for the tail of the distribution of $\ln{\tau}$.

\begin{figure}[tbhf]
\centering
\includegraphics[width=0.3\textwidth,angle=270]{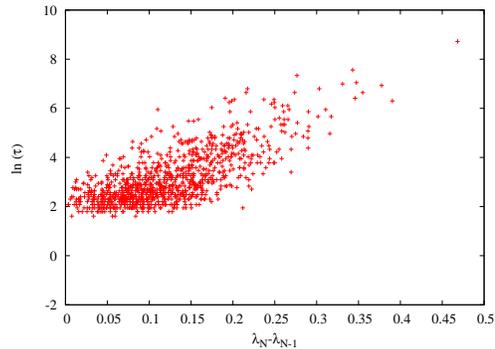}
\caption{The logarithm of the relaxation time as a function of the gap
  between the two largest eigenvalues of the $\{ J_{i,j} \}$ matrix
  for $N=64$ and $T=0.8$. There are no disorder samples with
  $\tau>t_{window}$ in the data.}
\label{eigendelta}
\end{figure}

In the spherical SK model~\cite{RODMOO} the relaxation time is fixed
by the difference between the two largest eigenvalues of the $\{
J_{i,j} \}$ matrix, through the equation

\begin{equation}
\ln{\tau}=N/(2T)(1-T)( \lambda_N-\lambda_{N-1})\ .
\label{MM}
\end{equation}
For the SK model with binary couplings we consider here, the
correlation between $\ln{\tau_Ï}$ and $\lambda_N-\lambda_{N-1}$ is
more fuzzy, as one can see in figure~\ref{eigendelta}. It is in fact
fuzzier than the relation between $\ln{\tau_Ï}$ and $\lambda_N$ alone.
This is true for all values of $N$ and $T$ considered, and for all
values of $\ln{\tau}$ (both large and small).  Note that in the
spherical SK model, equation~\ref{MM} is used to prove that
$E(\ln{\tau})\propto N^{1/3}$.  It turns out~\cite{ABEM} that this
scaling holds also in the usual SK model although equation~\ref{MM}
does not hold.

\begin{figure}[tbhf]
\centering
\includegraphics[width=0.3\textwidth,angle=270]{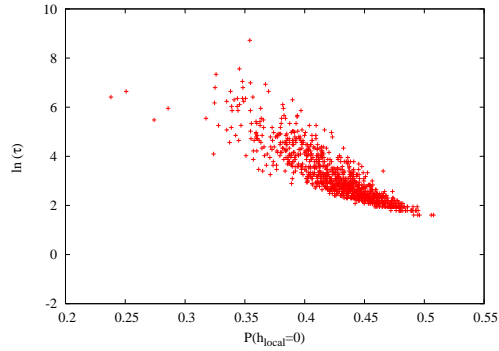}
\caption{The logarithm of the relaxation time as a function of
  $P(h_{local}=0)$ for $N=64$ and $T=0.8$. The normalisation is such that 
  $\int_0^{\infty} dh_{local} P(h_{local})=1$.
  There are no disorder
  samples with $\tau>t_{window}$ in the data.}
\label{pdh}
\end{figure}

In the spirit of~\cite{MONGAR}, we have also looked at the correlation
between $\ln{\tau}$ and the site averaged local field probability
distribution $P(h_{local})$

\begin{equation}
P(h_{local}) =\frac{1}{N} ( < \sum_i \delta ( h -\sum_{j\neq
  i}\ J_{i,j} \sigma_j ) > )\ .
\end{equation}
Specifically we looked at the correlation between $\ln{\tau}$ and the
value of $P_(h_{local})$ at the origin. Our results can be found in
figure~\ref{pdh} for $N=64$ and $T=0.8$.  In~\cite{MONGAR} it was
argued that the disorder samples with large relaxation times are the
ones with all local fields (on every site $i$) large.  This is
indicative of a correlation between large values of $\ln{\tau}$ and a
distribution $P_(h_{local})$ that is depleted at the origin.
(Indeed~\cite{TTCSS} $\ln{\tau}$ increases as $T$ decreases, while
$P(h_{local}=0)$ decreases as $T$ decreases, with
$P(h_{local}=0)=0$ at zero temperature in the $N=\infty$ limit).
Our data indeed show a neat correlation between large values of
$P(h_{local}=0)$, and small $\ln{\tau}$, but this correlation
becomes fuzzy for lower values of $P(h_{local}=0)$. Again we were
looking for a characterisation of the slow disorder samples but we
found a characterisation of the fast disorder samples instead. The
observed correlation seems to persists as $N$ grows, but fades  as $T$ is
decreased. There is a similar correlation between $\ln{\tau}$ and the
average value of $h_{local}$ within the tail of $P(h_{local})$, for
example within the last five percents of the distribution. A small
value of $h_{local}$ inside the tail is correlated to a small
relaxation time.

The strongest correlation we found is between $\ln{\tau}$ and the
average overlap squared $<q^2>$, as can be seen in figure~\ref{Q2}.
The observed correlation
seems to persists as $N$ grows, and as $T$ is decreased.
This correlation is an
empirical finding and we have no dynamical explanation for
it.  On the other hand $<q^2>$ is fairly easy to
estimate with Monte Carlo, even with little statistics and this
correlation could be used to flag slow samples in numerical
simulations.  We finally remark that if there is a correlation between
$\ln{\tau}$ and the average overlap squared $<q^2>$, there is no
correlation between $\ln{\tau}$ and the number of peaks in the order
parameter distribution $P(q)$.  We have looked for such a
correlation using the data of~\cite{ABMM}, where the number of peaks was
estimated for a subset of the disorder samples considered here.  The
presence of thermal noise in the measured distribution makes it
difficult to count the number of peaks.  In this reference the authors
did their best to count the number of peaks by visual inspection of
the plots of $P(q)$ for $192$ disorder samples, with $T=0.4$. We
find no correlation between the value of $\ln{\tau}$ and the number
of peaks for $N=64$ and $256$.

\begin{figure}[tbhf]
\centering \includegraphics[width=0.3\textwidth,angle=270]{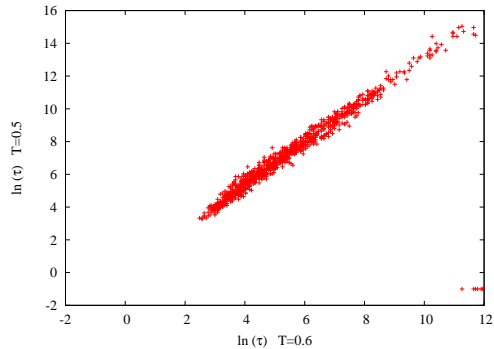}
\caption{Logarithm of the relaxation time for $T=0.5$ as a function of
  the logarithm of the relaxation time for the same disorder sample
  $J$ but $T=0.6$. The number of sites is $N=64$. There are a couple of
  disorder samples with $\tau>t_{window}$ for $T=0.5$ in this figure.
  Those are the points with $\ln{\tau}=-1$ (by convention).}
\label{CC}
\end{figure}

\begin{figure}[tbhf]
\centering \includegraphics[width=0.3\textwidth,angle=270]{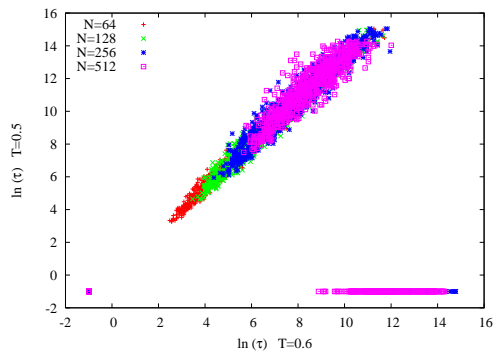}
\caption{(Colour on line) Same as in figure~\ref{CC} but with $N=64,
  128, 256$ and $512$ together. There are a couple of disorder samples
  with $\tau>t_{window}$ for both $T=0.5$ and $T=0.6$ in this figure.
  Those are the points with $\ln{\tau}=-1$ (by convention).}
\label{CCALL}
\end{figure}

We have found finally that, disorder sample by disorder sample, the
relaxation time measured at two temperatures (both in the spin glass
phase) are strongly correlated. This can be seen in figure~\ref{CC}
where we show $\ln{\tau(T=0.5)}$ as a function of $\ln{\tau(T=0.6)}$
both for $N=64$.  This figure shows a very strong correlation.  One
notice a couple of samples for which $\ln{\tau}(T=0.5)=-1$ (censored
data).  Figure~\ref{CCALL} shows the same figure but with all systems
sizes together. Obviously there are more and more disorder samples
censored as $N$ grows, with now a whole horizontal segment with
$\ln{\tau}=-1$.  Interestingly all points in the scatter plot scale on
the same $N$ independent thick line. This thick line would extend
further towards large values had we used a larger observation window.
(But it would be quite CPU time consuming to obtain a large
extension). Note that the $N=512$ data appear to be more scattered
than the other data, an optimistic interpretation is that we are
seeing some onset of a temperature chaotic behaviour. In order to be
more quantitative, we have analysed the data as follows: first we made
a linear least squares fit of the $N=64$ data to the form
$\ln{\tau}(T=0.5)=a+b \ln{\tau}(T=0.6)$, with parameters $a$ and $b$ 
(we obtain the values $a=0.195$ and $b=1.297$).
Then we computed, for every system size $N$, the deviation
$\delta_N^2=1/(N_J (1+b^2)) \sum_J \bigl(a+b
\ln{\tau}(T=0.6)-\ln{\tau}(T=0.5)\bigr)^2$, with the values of $a$ and
$b$ obtained from the fit of the $N=64$ data, an a sum over those
disorder samples such that both relaxation times are less than
$t_{window}(N/512)^{1/3}$, with $t_{window}$ the cutoff used for the
$N=512$ data, and $N_J$ the number of disorder samples that satisfy
the constraint. In words, $\delta_N^2$ is the mean squared deviation
from the linear squares fit, taking properly into account the
relaxation time observational cutoff.  We find that $\delta_N/N^{1/3}=
0.0423, 0.0485, 0.0509$ and $0.0568$ for $N=64, 128, 256$ and $512$
respectively. The correlation between the values of $\ln{\tau}$ at two
temperatures becomes looser as $N$ grows, confirming quantitatively
the indication of the onset of temperature chaos.

In conclusion, we have measured the equilibrium relaxation time $\tau$
of the Sherrington-Kirkpatrick model with binary couplings for many
samples of the quenched disorder, and several values of the
temperature, with system sizes from $N=64$ to $512$, taking great care
that the thermal noise is negligible.  We confirm the result
of~\cite{MONGAR} that the slow samples are not correlated to
``ferromagnetic" disorder configurations, but we did not find evidence
for a dominance by configurations with a small value at the origin of
the site averaged local field probability distribution. We find a
strong correlation between the relaxation times measured at two
distinct values of the temperature (with the same disorder sample).
Closer look shows a broadening as $N$ grows, that is possibly an indication of the
onset of temperature chaos.

\ack
I thank Jean-Philippe Bouchaud, Thomas Garel, Enzo Marinari and
C\'ecile Monthus for discussions. The numerical simulation were done
using the CCRT computer center in Bruy\`eres-le-Ch\^atel.

\end{document}